\documentclass[aps, prb,preprint]{revtex4-1}
\usepackage{amsmath, amsthm}
\usepackage{amssymb}
\usepackage{tikz}
\usepackage{lipsum}
\usepackage{amsfonts}
\usepackage{setspace}
\usepackage{amssymb}
\usepackage{graphicx}
\usepackage{color}
\usepackage{appendix}
\usepackage[justification=justified,singlelinecheck=false]{caption}
\usepackage{subfigure}
\usepackage{afterpage}
\usepackage{dcolumn}
\usepackage{ragged2e}
\usepackage{url}
\usepackage{multirow}
\usepackage{enumerate}
\usepackage{epstopdf}
\usepackage[margin=2cm]{geometry}
\begin{document}\title{Reinterpretation of bond-valence model with bond-order formalism: an improved bond-valence based interatomic potential for PbTiO$_3$ }
\date{\today}
\author{ Shi Liu, Ilya Grinberg, Hiroyuki Takenaka, and Andrew M. Rappe}
\affiliation{The Makineni Theoretical Laboratories, Department of Chemistry,University of Pennsylvania, Philadelphia, PA, 19104-6323 }
\begin{abstract}
We present a modified bond-valence model of PbTiO$_3$ based on the principles of  bond-valence and bond-valence vector conservation. The relationship between the bond-valence model and the bond-order potential is derived analytically in the framework of a tight-binding model. A new energy term, bond-valence vector energy, is introduced into the atomistic model and the potential parameters are re-optimized. The new model potential can be applied both to canonical ensemble ($NVT$) and isobaric-isothermal ensemble ($NPT$) molecular dynamics (MD) simulations. This model reproduces the experimental phase transition in $NVT$ MD simulations and also exhibits the experimental sequence of temperature-driven and pressure-driven phase transitions in $NPT$ simulations. We expect that this improved bond-valence model can be applied to a broad range of inorganic materials. 
\end{abstract}

\maketitle
The use of ferroelectric perovskite oxides in a variety of technological applications has prompted extensive investigations of their structure and dynamics.~\cite{Lines77,Scott07p954} First-principles density functional theory (DFT) calculations  have played an important role in enhancing microscopic understanding of the relationships between composition, structure and properties.~\cite{Zhong94p3618, Cohen92, Grinberg04} Despite the success of first-principles methods, the great computational expense and the difficulties of studying finite-temperature properties have driven the development of more efficient atomistic and effective Hamiltonian potentials suitable for large-scale molecular dynamics (MD) simulations.~\cite {Pettifor99,Sepliarsky02,Sepliarsky05,Grinberg02,Pettifor02,Shin05,vanDuin05,Aoki07,Shimada08,Zhong95p6301,Waghmare97p6161,Wojdel13} In particular, an atomistic potential based on the widely used bond-valence (BV) theory~\cite{Brown09} has been developed.~\cite{Grinberg02, Shin05}  BV-based atomistic potentials have since been used to study phase transitions~\cite{Shin08p015224} and domain wall motion in PbTiO$_3$~\cite{Shin07}, as well as structure and dynamics in the classic 0.75PbMg$_{1/3}$Nb$_{2/3}$O$_3$-0.25PbTiO$_3$ relaxor ferroelectric material.~\cite{Grinberg09,Takenaka13}\\
\indent The bond-valence theory, or bond-valence conservation principle, states that in a crystal structure, each atom $i$ prefers to obtain a certain atomic valence, $V_{0,i}$. The actual atomic valence $V_i$, for atom $i$ can be obtained by summing over the bond valences $V_{ij}$, which can be calculated from an empirical  inverse power-law relationship~\cite{Brown73,Brown76} between bond valence and bond length $r_{ij}$:
\begin{equation}
V_{ij}=\left(\frac{r_{0,ij}}{r_{ij}}\right)^{C_{ij}}.
\end{equation}
$r_{0,ij}$ and $C_{ij}$ are Brown's empirical parameters. 
The energy contribution of the bond-valence is chosen to have the following form:
\begin{equation}
\label{BVE}
E_{BV}=\sum_i\varepsilon_{i}=\sum_i{S_i(V_i-V_{0,i})}^2,
\end{equation}
where $\varepsilon_{i}$ is the atomic bond-valence energy and $S_i$ is a scaling parameter. \\
\indent Despite the success of the rather simple ten-parameter BV model potential~\cite{Shin05} for PbTiO$_3$, no rigorous quantum mechanical justification has been provided for the bond-valence potential energy, raising questions about the general applicability of this type of atomistic potential. In addition, the potentials obtained in previous work~\cite{Shin05,Shin08p015224} were found to be accurate for $NVT$ simulations only, with incorrect ground state structures obtained when the constant volume constraint is lifted.  In this paper, we show how the bond-valence energy can be derived from the second-moment bond-order potential, extend the model to represent higher moments of the local density of states (LDOS), and show that this allows accurate simulations for both constant-volume and constant-pressure conditions.\\
\indent An analysis of the physics that gives rises to the bond-valence conservation principle shows that the bond-valence energy can be naturally derived from the second-moment bond-order potential, such as the well-known Finnis-Sinclair potential.~\cite{Finnis, Sutton04} Within the framework of a tight-binding model~\cite{Horsfield96}, the Finnis-Sinclair potential can be partitioned into atomic contributions as:
\begin{equation}
U_{\rm FS}(\mathbf{r}_{1}, ...\mathbf{r}_{\rm N})=\sum_i E_i=\sum_i\left[\sum_{\left<j\right>}\phi(r_{ij})-\gamma_i(\mu_i^{(2)})^{\frac{1}{2}}\right],
\end{equation}
\noindent where $\mathbf{r}_{i}$ is the atomic position, $E_i$ is the local atomic energy and $\phi(r_{ij})$ is a pair-wise repulsive potential depending on the distance between atom $i$ and its nearest-neighboring atom $j$. The second term represents the bonding energy; $\gamma_i$ is a constant and $\mu_i^{(2)}$ is the second moment of the LDOS. The second moment $\mu_i^{(2)}$ measures the width of the LDOS distribution, and as shown by Cyrot-Lackmann and Ducastelle,~\cite{Cyrot68p1235,Cyrot90p2744,Ducastellep285} can be evaluated from the summation over all the nearest-neighbor hopping paths that start and end on atom $i$:
\begin{equation}
\mu_i^{(2)}=\sum_{\left<j\right>}\beta_{ij}\beta_{ji}=\sum_{\left<j\right>}\beta_{ij}^2,
\end{equation}
\noindent where $\left<j\right>$ means the summation of nearest neighbors of $i$, and $\beta_{ij}$ is the {\em averaged} hopping integral between atom $i$ and $j$. Because the overlap of atomic orbitals decays as $\exp(-\sigma_{ij}r_{ij})$~\cite{Sutton04} , Eq~(3) can be written as 
\begin{equation}
\label{UFS}
U_{\rm FS}=\sum_i E_i=\sum_i \sum_{\left<j\right>}a_{ij}e^{-2\sigma_{ij}r_{ij}}
-\sum_i\gamma_{i}\left(\sum_{\left<j\right>}b_{ij}e^{-2\sigma_{ij}r_{ij}}\right)^{\frac{1}{2}} ,
\end{equation}
\noindent with $\phi(r_{ij})=a_{ij}e^{-2\sigma_{ij}r_{ij}}$ and $\mu_{i}^{(2)}=\sum_{\left<j\right>}b_{ij}e^{-2\sigma_{ij}r_{ij}}$, where $a_{ij}$ is a constant that scales the strength of the repulsive interactions between atom $i$ and atom $j$, and $b_{ij}$ scales the bonding interaction.  \\
\indent Despite the different appearance of Eq~({\ref{BVE}}) and Eq~({\ref{UFS}}), we can rewrite the bond-valence energy in a similar form to the FS potential. First of all, we point out that the energy function for bond-valence energy is not unique since Eq~({\ref{BVE}}) simply enforces that any deviation from the desired atomic valence will incur an energy penalty.  In principle, any energy function that reflects this principle should be equivalent to Eq~({\ref{BVE}}). Therefore, we could rewrite the bond-valence energy as
\begin{equation}
\label{BVE2}
E_{BV}=\sum_i{S'_i(\sqrt{V_i}-\sqrt{V_{0,i}})}^2,
\end{equation}
with $S'$ as a scaling parameter. The bond-valence is an empirical concept, and it has been modeled with various functional forms including inverse power law and exponential.~\cite{Brown09} For the narrow range of distances of first nearest neighbor pairs, exponential and power law yield similar results. Given that the bond-valence reflects the bonding strength, we define it as an exponential of the interatomic distance: 
 \begin{equation}
\label{BV2}
V_{ij}=b_{ij}'e^{{-2\sigma_{ij}r_{ij}}}
\end{equation} 
where $b_{ij}'$ is a parameter depending upon the type of atomic pair.  
Expanding Eq~({\ref{BVE2}}) gives
\begin{equation}
\label{BVE3}
E_{BV}=\sum_i{S'_iV_{i}-2S'_i\sqrt{V_{0,i}V_{i}}+S'_iV_{0,i}}.
\end{equation}
The last term, $S'_iV_{0,i}$, is a constant and will cancel out when  energy differences are considered. Henceforth we will not write out this constant term explicitly. 
Substituting Eq~({\ref{BV2}}) into Eq~({\ref{BVE3}}), we obtain 
\begin{equation}
\label{BVE4}
E_{BV}=\sum_i\sum_{\left<j\right>}S'_ib_{ij}'e^{-2\sigma_{ij}r_{ij}}-\sum_{i}2S'_i\sqrt{V_{0,i}}\left(\sum_{\left<j\right>}b_{ij}'e^{-2\sigma_{ij}r_{ij}}\right)^{\frac{1}{2}}.
\end{equation}
It becomes evident that the bond-valence energy expressed in Eq~({\ref{BVE4}}) is remarkably similar to the FS potential in Eq~({\ref{UFS}}). Eq~({\ref{UFS}}) and Eq~({\ref{BVE4}}) becomes equivalent if we choose
\begin{subequations}
\label{BV5}
\begin{equation}
S_i' b_{ij}' = a_{ij}
\end{equation}
\begin{equation}
2S_i' b_{ij}' \sqrt{V_{0,i}} = b_{ij}
\end{equation}
\end{subequations}
Rearranging Eq~(\ref{BV5}), we obtain $V_{0,i} = {b_{ij}^2}/{4a_{ij}^2}$. Therefore, for any system where the ratio of coefficients for bonding and repulsive interactions, $b_{ij}/a_{ij}$, is constant among the neighbors of atom $i$, this ratio defines this atom's bond valence.  Thus, the bond valence energy Eq~({\ref{BVE4}}) is equivalent to  Eq~({\ref{UFS}}).  The equivalence between the bond-valence energy and the Finnis-Sinclair potential means that the bond-valence conservation experimentally observed in solids is based on the quantum-mechanical description of bonding that underlies the Finnis-Sinclair model. \\ 
\indent Compared to the bond-order potential, the application of the bond-valence model does not require extra efforts to parametrize hopping integrals, because the bond-valence parameters for a wide variety of atomic pairs are already known from crystallography.~\cite{Brown09} Since the bond-valence model is a second-moment bond-order potential, its limitations, such as the inability to obtain the correct ground state structure in $NPT$ simulations, are likely due to the fact that the second moment only accounts for the width of LDOS but does not reflect its shape. One consequence of this is that the BV energy depends only on the total valence and is entirely insensitive to the number of bonds or their relative strengths. This feature of all second-moment models makes it difficult to distinguish between competing crystal structures, which are controlled by the higher moments.~\cite{Sutton04} Therefore, a systematic way to improve the bond-valence model is to include the contributions of higher moments of the LDOS (such as fourth moment) to the total energy.~\cite{Carlsson83, Hansen91}\\

In this work, we choose the bond-valence vector sum (BVVS)~\cite{Harvey06,Brown09} to reflect the change of the fourth moment of the LDOS. The bond-valence vector is defined as a vector lying along the bond with magnitude equal to the bond-valence ($|\mathbf{V}_{ij}|=V_{ij}$), as shown in Figure 1. A simple argument is presented in the Appendix to illustrate the relationship between the fourth moment of the LDOS and the sum of the bond-valence vectors in a periodic structure. Generally, the changes in the local symmetry of the bonding environment affect the value of the fourth moment of the LDOS, which is also reflected by the change of BVVS. We suggest that BVVS is a natural way to capture the change in the fourth moment of LDOS. For many materials, it has been shown that the ground-state structure favors symmetric local bonding environment and a zero BVVS. Therefore, the criterion of BVVS = 0 for the ground-state structure has been suggested as a complement to the original bond-valence conservation principle.~\cite{Harvey06,Brown09}  However, this is not followed for crystal structures in which symmetry breaking (BVVS $\neq$ 0) becomes significant due to electronic-structure driven distortions, such as the second order Jahn-Teller distortion exhibited by Ti atoms in an octahedral environment and the stereochemical lone-pair driven distortions of Pb$^{2+}$ cation. The BVVS can thus be considered as a measure of local symmetry breaking. We therefore generalize this principle by proposing that each ion has a desired length of bond-valence vector sum. The bond-valence vector energy, $E_{BVV}$, is defined as
\begin{equation}
E_{BVV}=\sum_i{D_i(\mathbf{W}_i^2-\mathbf{W}_{0,i}^2)}^2,
\end{equation}
where 
\begin{equation}
\mathbf{W}_i=\sum_{j\ne i}\mathbf{V}_{ij}=\sum_{j\ne i}V_{ij}\mathbf{{\hat R}}_{ij}.
\end{equation}
$D_i$ is the scaling factor, $\mathbf{W}_i$ is the calculated bond-valence vector sum and $\mathbf{W}_{0,i}$ is the desired value of bond-valence vector sum. It is noted that only the norm of the bond-valence vector sum is taken in the energy term (square of $\mathbf{W}_{i}$) since the energy is a scalar quantity and the energy expression should bot break the system symmetry. The value of $\mathbf{W}_{0,i}$ can be computed using the optimized atomic positions in the lowest-energy structure identified from first principles. We note here that the proposed BVV energy is a simplified fourth-moment bond-order potential, as the calculation of BVVS for a given atom only requires the knowledge of its nearest neighbors. \\
\indent The interatomic potential for our modified bond-valence model is given by:
\begin{equation}
E=E_c+E_r+E_{BV}+E_{BVV}+E_a
\end{equation}
\begin{equation}
E_c=\sum_{i<j}\frac{q_iq_j}{r_{ij}}
\end{equation}
\begin{equation}
E_r=\sum_{i<j}\left(\frac{B_{ij}}{r_{ij}}\right)^{12}
\end{equation}
\begin{equation}
E_a=k\sum_{i}^{N_{\rm oxygen}}(\theta_{i}-180^{\circ})^2
\end{equation}
\noindent where $E_c$ is the Coulomb energy and $E_r$ is the short-range repulsive Lennard-Jones energy. In both the Finnis-Sinclair potential and the bond-valence model, only averaged hopping integrals between neighboring atoms are used, which is equivalent to approximating all the atomic orbitals as $s$-type.~\cite{Sutton04} However, bonding in PbTiO$_3$ involves $p$-$d$ orbital hybridizations, which do display angular dependence. Physically, in PbTiO$_3$ this results in an energy cost for rotations of oxygen octahedra. To introduce the dependence of energy on the interatomic angles, we include an angle potential term, $E_a$, which is defined locally for all the O-O-O angles along the oxygen octahedral axes, as shown in Figure 2. This rotationally-invariant angle potential prevents unphysically large tilting of oxygen octahedra. \\
 \indent The potential parameters required to be fitted for PbTiO$_3$ can be summarized as follows: spring constant $k$ for angle potential, charges $q_{i}$, scaling factors $S_i$ and $D_i$ for each species, and short-range repulsion parameters, $B_{ij}$, for each pair type (Pb-Ti, Pb-O, Ti-O and O-O). The Brown's empirical parameters ($r_{0,ij}$ and $C_{0,ij}$) are taken from Ref. 23 and Ref. 24. We implemented this bond-valence model in the LAMMPS code.~\cite{Plimpton}\\
\begin{table*}[t]
    \caption{{Optimized potential parameters of modified bond-valence model. The angle potential parameter $k$ is 0.0152 eV/(deg)$^2$}. }
    \label{Table1}
      \begin{center}
      \begin{tabular}{ccccccc|ccc|cc}  
     \hline\hline
  &		&&&		&		&		&	&$B_{\beta \beta '} $(\AA)	&	&	&\\
  \hline
  &		&$r_{0,\beta \rm O}$&$C_{0,\beta \rm O}$&	$q_{\beta}$(e)	&	$S_{\beta}$(eV)	& $D_{\beta}$	&Pb    &	Ti		&	O	&	$V_{0,\beta}$ &$\mathbf{W}_{0,\beta}$\\
  \hline
   &	Pb	&1.960&5.5&	1.38177	&	0.31646	&	2.23180	&	-- &2.17558	&  1.71871               &2.00	&     0.40297	\\
   &	Ti	&1.798&5.2&	0.99997 &	--	       &	0.11888	&	-- &		--	&1.28582	&4.00	&	0.46541\\
   &	O	&-&-&	-0.79391	&	1.52613	&	--	          &	-- &		--	&1.83109	&2.00	&	--\\
  \hline \hline
      \end{tabular}
    \end{center}
  \end{table*}  
 \indent Figure  3 shows our parameterization protocol. The optimization of the potential parameters is performed using simulated annealing (SA) global optimization method to fit a database of structural energy differences and atomic forces ($E$ \& {\bf F}) derived from {\em ab initio} DFT calculations with the ABINIT code.~\cite{Gonze02} We used the $2\times2\times2$ supercell as the reference structure. The energy and atomic forces are computed with $2\times2\times2$ Monkhorst-Pack $k$-point mesh~\cite{Monkhorst76} using PBEsol ~\cite{Perdew08} as the exchange-correlation energy functional. We start with an initial database that contains the lowest-energy tetragonal structure, strained tetragonal structures, the lowest-energy cubic structure, strained cubic structures, and randomly picked orthorhombic structures with various lattice constants. After each SA run, the optimized potential parameters are used to perform constant-stress MD simulations to generate equilibrium structures at various temperatures, which are then put back to the database. The process is continued until the energies and forces of the structures sampled during MD simulations are accurately reproduced (difference between MD value and DFT value is $\approx$4~meV/atom). \\
\indent Table I presents the optimized potential parameters. To account for the overestimation of the  PbTiO$_3$ $c/a$ ratio  by  PBEsol ($c/a$=1.10 versus $c/a$=1.07 experimentally)~\cite{Zhao08}, we adjusted Brown's empirical parameter $r_{0,ij}$  to make the $V_{\beta}$ for Pb, Ti and O reach their atomic valences in the lowest-energy tetragonal structure obtained with PBEsol.  The value of preferred BVVS is then calculated with the modified $r_{0,ij}$. We find that the oxygen atoms do not have a preference for a specific value of bond-valence vector sum. This is because in perovskites, some oxygen atoms are highly displaced ($|\mathbf{W}_{\rm O}| > 0$), while others stay around the high-symmetry point ($|\mathbf{W}_{\rm O}| = 0$). So the BVVS term is included for Pb and Ti only.\\
\indent  Using this optimized model potential for PbTiO$_3$, we studied the temperature dependence of lattice constants, polarization and displacements of Pb and Ti ions using an $8\times8\times8$ supercell. We first performed canonical-ensemble MD simulations with lattice constants fixed to experimental values, using the Nos\'{e}-Hoover thermostat to control the temperature. Figure 4(a) shows the evolution of polarization at different temperatures: only $P_z$ along the $c$ axis has significant values at low temperature and the overall polarization becomes zero at and above $T_c$. For these simulations, we obtained 830~K for the ferroelectric-to-paraelectric first-order phase transition temperature $T_c$, shown in Figure 4(b). This agrees well with the experimental $T_c$ of 765~K,~\cite{Shirane51} and is an improvement relative to the 550~K value obtained in $NVT$ calculations with an earlier BV potential without BVVS term.~\cite{Shin05, Shin08p015224}  We then used the new potential in $NPT$ simulations, with the pressure maintained at 0.1 MPa by the Parrinello-Rahman barostat.~\cite{Parrinello80} For the  ground state structure at 10~K, we obtained the lattice constant  $a$=3.834~\AA\ and $c/a$=1.15. The equilibrium $c$/$a$ ratio in MD is larger than the PBEsol DFT value.~\cite{Zhao08} Figure 5 displays the temperature dependence of lattice constants, spontaneous polarization and atomic displacements of Pb and Ti obtained from $NPT$ simulations. As temperature increases, the $c/a$ ratio decreases gradually, together with the polarization and atomic displacements. The phase transition from tetragonal to cubic occurs at 400~K, lower than the experimental value. The rather large magnitude of spontaneous polarization compared to experimental value ($P$ = 1.25~C/m$^2$ vs. experimental $P$ = 0.75~C/m$^2$)~\cite{Gavrilyachenko} and the large atomic displacements at temperatures below $T_c$ are due to the overestimated tetragonality of the PBEsol functional and some amplification of this effect in the resulting potential. \\
\indent We find that the new potential is capable of describing domain wall (DW) energetics and structures. The supercell used to model the domain wall is constructed following the method in Ref. 42. The domain wall energy ($E_{\rm DW}$) is calculated by
\begin{equation}
E_{\rm DW}=\frac{E_{N}-E_{\rm bulk}}{S_{\rm DW}},
\end{equation}
where  $E_{N}$ is the energy of the supercell, $E_{\rm bulk}$ is the energy of a single-domain supercell of the same size, and $S_{\rm DW}$ is the area of the domain wall.  Figure 6(a) presents simulation of 180$^\circ$ Pb-centered domain walls at 10~K.  The computed domain wall energy is 208 mJ/m$^2$, agreeing very well with 170 mJ/m$^2$ obtained via PBEsol DFT calculations (with an $8\times1\times1$ supercell). To simulate a 90$^\circ$ domain wall, we used a supercell with $N_1=16$, $N_2=4$, and $N_3=4$, as shown in Figure 6(b). The dimensions of the supercell are fixed to the values calculated based on experimental lattice constants of tetragonal PbTiO$_3$. The domain wall energy is estimated to be 90 mJ/m$^{2}$ and also shows a satisfying agreement with the PBEsol DFT value of 64 mJ/m$^2$ (with an $8\times1\times1$ supercell). We note that the BV potential is highly efficient, as all the interactions are pair-wise. This allows simulation of  a 40$\times$40$\times$40 supercell (320,000 atoms) for 40 ps with a 1.0 fs timestep using only 2268 seconds of clocktime with 320 CPUs on the iBM iDataPlex supercomputer at the Navy DoD Supercomputing Resource Center.\\
 \indent We have also  examined the performance of the potential in simulations of pressure-induced phase transitions in PbTiO$_3$ with a $10\times10\times10$ supercell.  Figure 7 shows the pressure dependence of lattice constants and polarization.  We find two phase transitions, at 6.5 GPa and 11 GPa. Below 6.5 GPa, the structure is ferroelectric. The tetragonality decreases with increased pressure and the magnitude of polarization along the long axis reduces accordingly. Above 6.5 GPa, the $c/a$ ratio becomes 1 but the structure maintains ferroelectricity up to 11~GPa. Between 6.5~GPa and 11~GPa, we find the coexistence of multiple monoclinic phases. The polarization disappears when the pressure exceeds 11~GPa and the structure becomes centrosymmetric and paraelectric. Our simulated results are consistent with Wu and Cohen's first-principles studies~\cite{Wu05,Ganesh09} and recent experimental results by Ahart {\em et al}.~\cite{Ahart08} We did not find any reentrance of ferroelectricity up to 60~GPa.\\
  
We have shown that bond-valence energy is formally equivalent to the second-moment bond-order potential. The introduction of bond-valence vector energy based on the bond-valence vector conservation principle improve the bond-valence model. The new potential of PbTiO$_3$ reproduces the polarization, ferroelectric instability and phase transition in $NVT$ simulations, and also captures the temperature-driven phase transition qualitatively in $NPT$ simulations. Both calculated 180$^\circ$ DW energy and  90$^\circ$ DW energy using this new potential are in agreement with DFT values. This new potential is efficient enough to simulate large supercells. The studies of pressure-induced phase transition with the new potential show two phase transitions, consistent with previous experimental studies. We expect that this improved bond-valence model can be applied to other oxides due to its simplicity, efficiency and accuracy.~\cite{Shi13}\\

S.L. was supported by the NSF through Grant CBET-0932786. H.T. was supported by the US DOE BES under Grant No. DE-FG02-07ER46431. I.G. was supported by the Energy Commercialization Institute. A.M.R. were supported by the US ONR under Grant No. N00014-11-1-0578. Computational support was provided by the Center for Piezoelectrics by Design, by the DoD HPCMO, and by the NERSC. We thank Tingting Qi for fruitful discussions. \\
\section*{Appendix}
The bond valence of an individual bond $V_{ij}$ is defined in Eq (7) to be proportional to the square of hopping integral $\beta_{ij}$. Both the bond-valence vector sum, $\mathbf{W}_{i} $, and the fourth-moment of the LDOS, $\mu_i^{(4)}$, can reflect the change of local symmetry of bonding environment. Figure A1 gives an example of a one-dimensional $AB$ alloy. The desired bond valence of $A-B$ in the undistorted structure is set to be $a$, and therefore the hopping integral is equal to $ \sqrt{\chi a}$, where $\chi$ is a constant. It is easy to calculate that the bond valence summation and $\mu^{(2)}$  at atom $A$ are $2a$ and $2\chi a$, respectively. Suppose that the lattice constant and A-B bond distances are changed such that the bond-valence of the longer $A-B$ bond to $(a-\delta)$ and the shorter one becomes $(a+\delta)$. Accordingly, the hopping integral for the longer $A-B$ become $\ \sqrt{\chi(a-\delta)}$ and the shorter one $\sqrt{\chi (a+\delta)}$.  The bond-valence conservation principle is obeyed in both structures so they cannot be distringuised at the second moment or bond-valence level. However, the $\mathbf{W}_{A} $ changes from zero in the undistorted structure to $2\delta$ in the distorted structure, and the $\mu_A^{(4)}$ is reduced from $6\chi^2 a^2$ to $6\chi^2 a^2-2\chi^2\delta^2$. It is evident that only the hopping path involving the next-nearest neighbors contributes to the change of fourth-moment. Since the fourth moment hopping terms and the BVVS change at the same order, the change of fourth moment, $\Delta \mu_{i}^{(4)}$, can be approximated with $(|\mathbf{W}_{i}|-|\mathbf{W}_{i,0}|)^2 $. We choose $\mathbf{W}_{i}^2$ instead of  $|\mathbf{W}_{i}|$ in the formula of $E_{BVV}$ to make sure $E_{BVV}$ is a differentiable function for each $\mathbf{W}_{i}$. \\
   
\newpage
\begin{figure}[!]
   \includegraphics[scale=1.0]{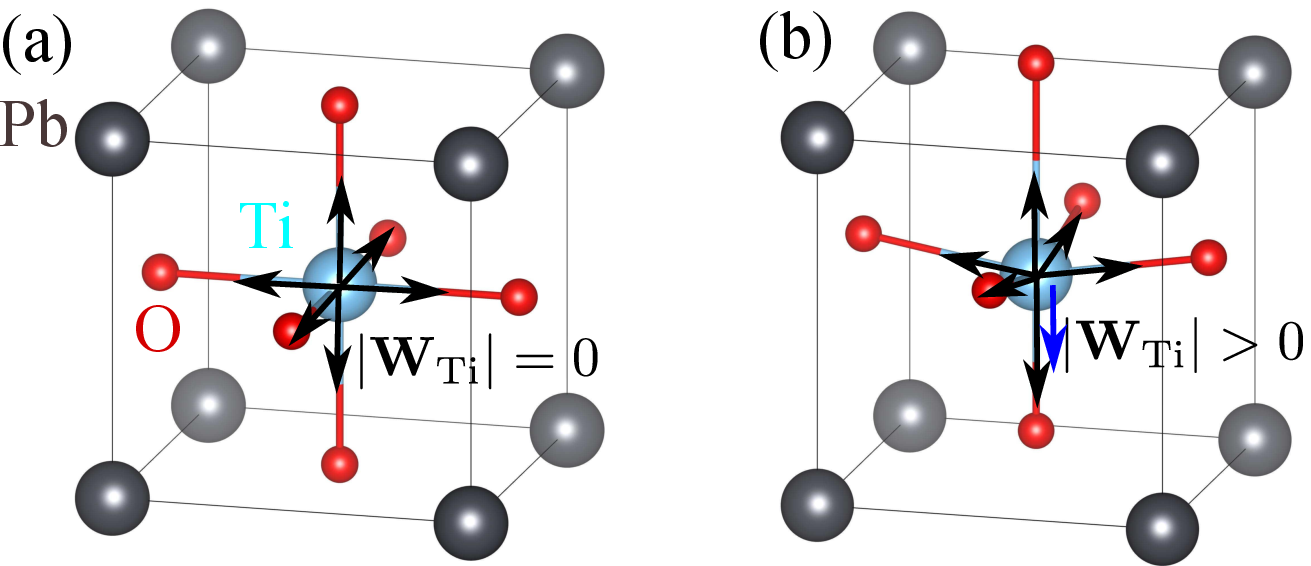} 
   \caption{(Color online) Schematic representation of bond-valence vector summation around Ti in (a) cubic PbTiO$_3$ and (b) tetragonal PbTiO$_3$. Gray, blue and red balls denote Pb, Ti and O. The back arrows scale the individual bond-valences, and the blue arrow shows the resultant bond-valence vector sum $\mathbf{W}_{\rm Ti}$.}
   \label{fig1}
\end{figure}
\begin{figure}[!] 
   \centering
   \includegraphics[scale=1.0]{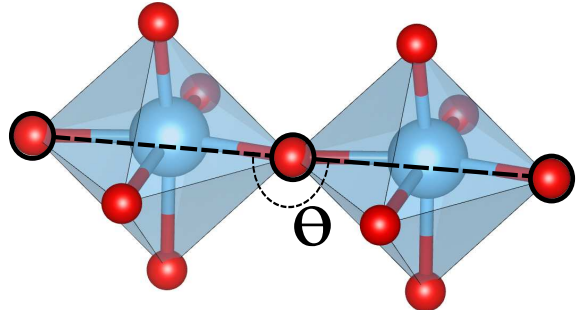} 
   \caption{Angle potential in bond-valence model.}
   \label{fig2}
   \end{figure}
 \begin{figure}[!] 
   \centering
   \includegraphics[scale=0.45]{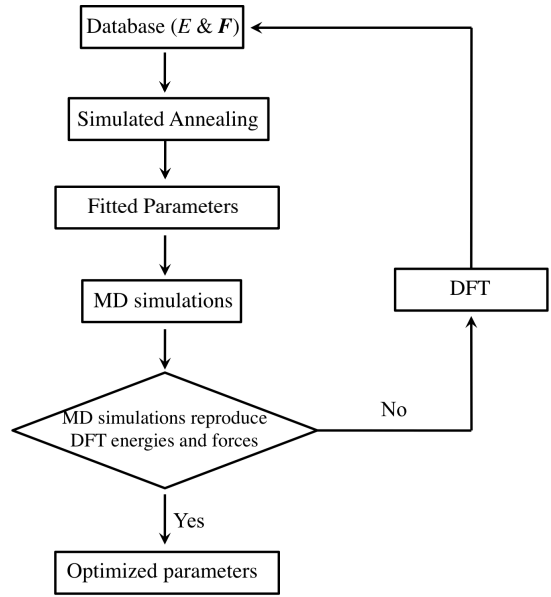} 
   \caption{Potential optimization protocol used in this work.}
   \label{fig3}
   \end{figure}
  \begin{figure}[!] 
   \centering
   \includegraphics[scale=1.5]{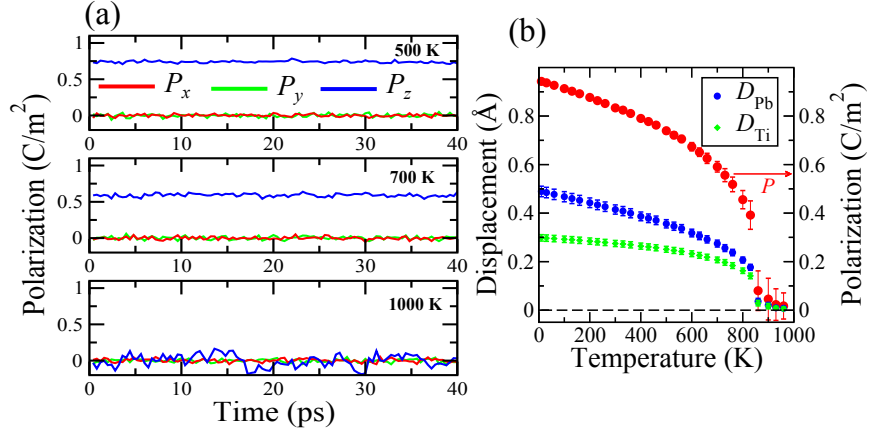} 
   \caption{(Color online) Temperature-dependent properties of PbTiO$_3$ obtained from $NVT$ simulations with lattice constants fixed to experimental values. The $c$ axis is along $z$ direction. (a) Time evolution of components of polarization for various temperatures. (b) Spontaneous polarization and atomic displacements along the $c$ axis as a function of temperature.}
   \label{fig4}
   \end{figure}   
  \begin{figure*}[!] 
  \centering
   \includegraphics[scale=2.00]{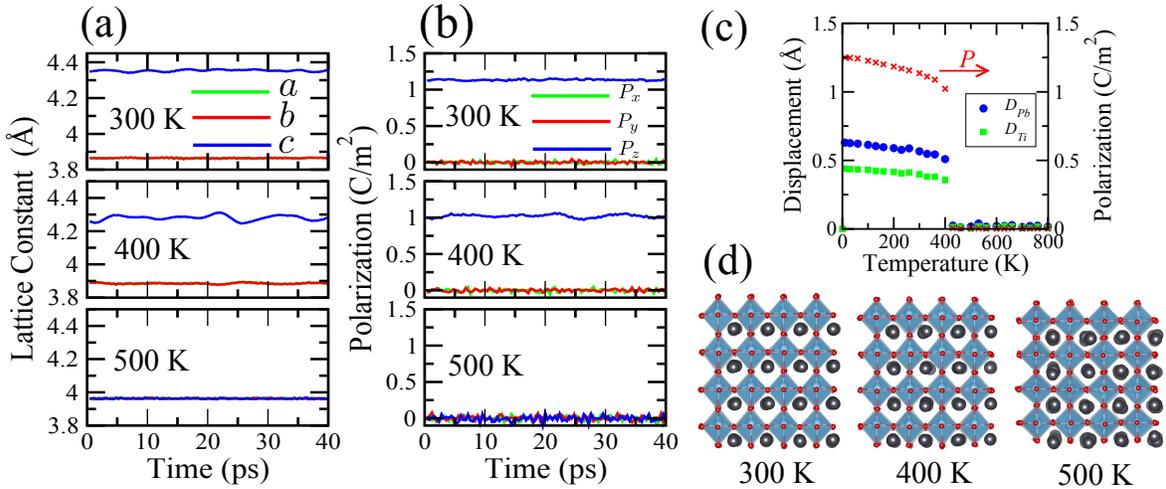} 
   \caption{(Color online) Temperature-dependent properties of PbTiO$_3$ obtained from $NPT$ simulations. Time dependence of (a) profiles of lattice constants and (b) profiles of polarization along the Cartesian axes for various temperatures. (c) Spontaneous polarization and atomic displacements as a function of temperature. (d) Snapshots of the structures of PbTiO$_3$.}
   \label{fig5}
  \end{figure*}  
\begin{figure}[!] 
   \centering
   \includegraphics[scale=1.5]{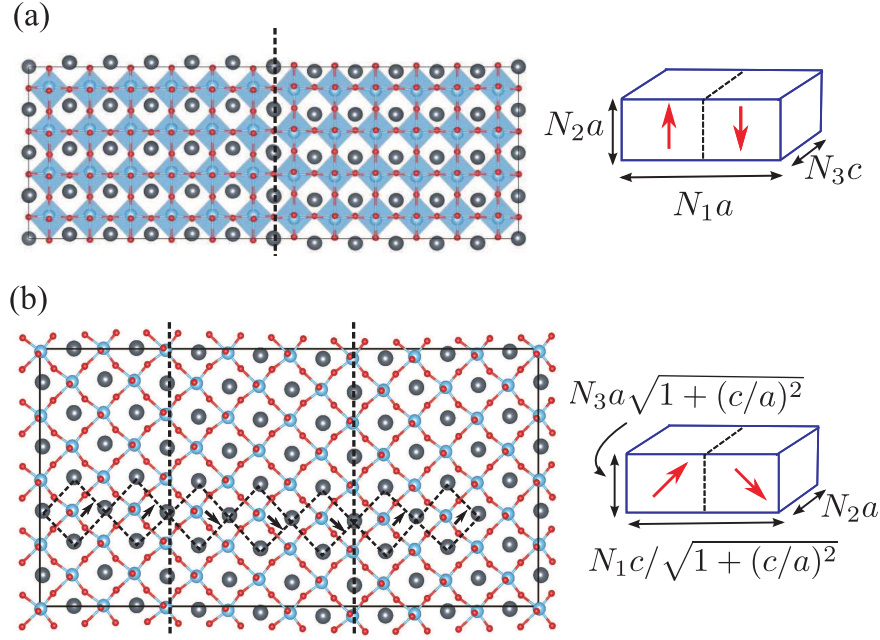} 
   \caption{(Color online) Simulated domain wall using modified bond-valence model. (a) 180$^\circ$ domain wall constructed with a 12$\times$4$\times$4 supercell; (b) 90$^\circ$ domain wall with $N_1=16$, $N_2=4$, $N_3=4$.}
   \label{fig6}
   \end{figure}
    \begin{figure}[!] 
   \centering
   \includegraphics[scale=0.5]{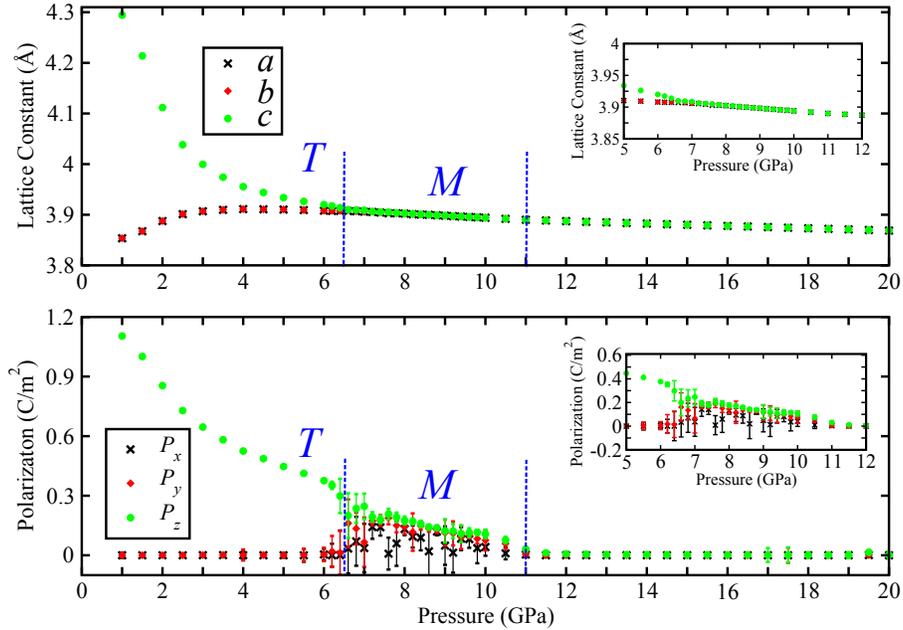} 
   \caption{(Color online) Pressure-induced phase transitions of PbTiO$_3$ obtained from MD simulations. Lattice axes coincide with the Cartesian axes ($a$ along $x$, $b$ along $y$ and $c$ along $z$).}
   \label{fig7}
   \end{figure}
   \begin{figure}[!] 
   \centering
   \includegraphics[scale=0.3]{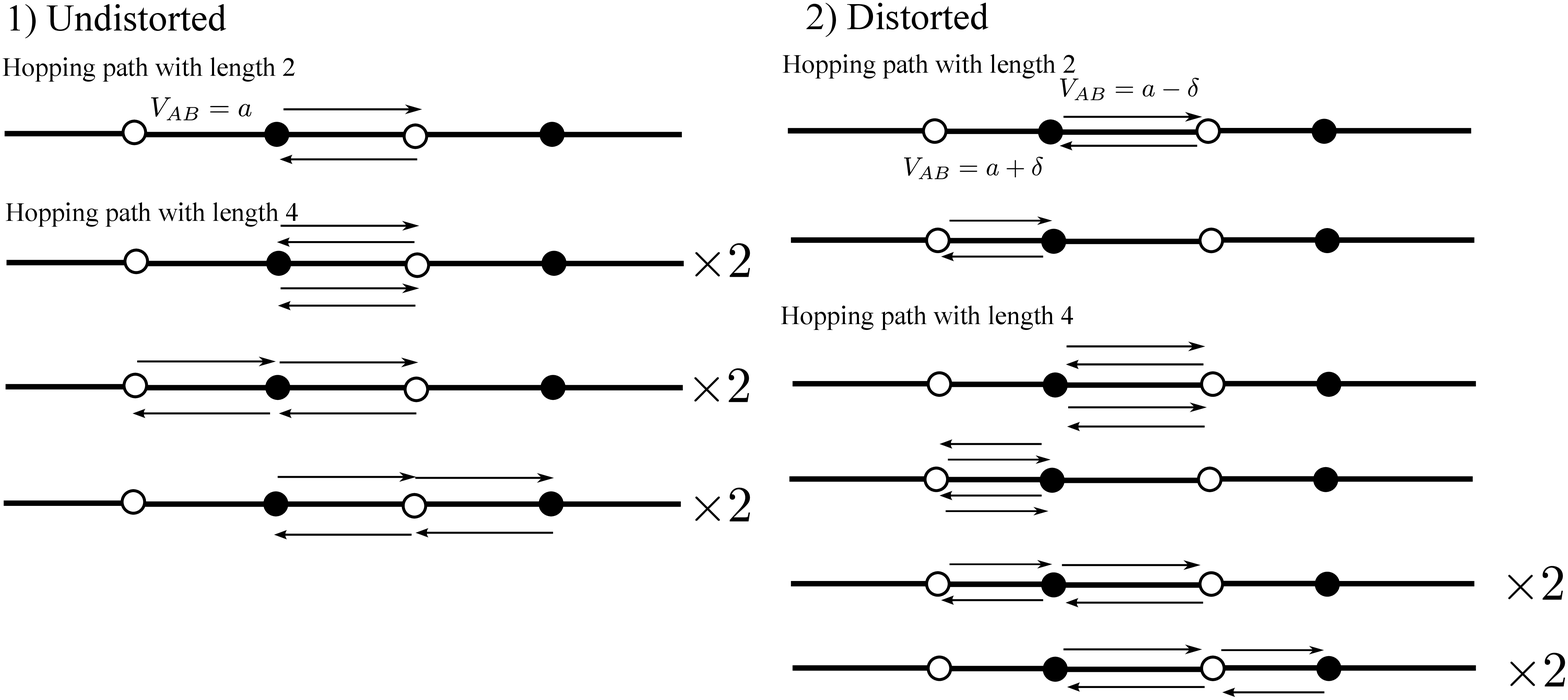} 
   \caption*{FIG. A1: Hopping paths in one-dimensional $AB$ alloy. Empty and filled circles represent elements $A$ and $B$. The bond-valence between $A$ and $B$ is represented as $a$. }
   \label{figA1}
   \end{figure}  

\end{document}